\documentclass[12pt,preprint]{aastex}

\usepackage{epsfig}
\usepackage{epsfig}
\usepackage{natbib}
\usepackage{lscape}
\usepackage{enumerate}
\usepackage{multirow}
\usepackage{array}

\def\plottwo#1#2{\centering \leavevmode
\includegraphics[width=.45\columnwidth]{#1} \hfil
\includegraphics[width=.45\columnwidth]{#2}}

\newcommand{\cN}[1]{\mathcal{N}}

\def\gsim{\;\rlap{\lower 2.5pt
 \hbox{$\sim$}}\raise 1.5pt\hbox{$>$}\;}
\def\lsim{\;\rlap{\lower 2.5pt
   \hbox{$\sim$}}\raise 1.5pt\hbox{$<$}\;}

\begin{document}


\title{%
Extreme Climate Variations\\
from Milankovitch-like Eccentricity Oscillations\\
in Extrasolar Planetary Systems\\
\begin{center}
{\footnotesize
Proceedings of the Milutin Milankovitch Anniversary Symposium in
2009:\\
\scriptsize Climate Change at the Eve of the Second Decade of the
Century}
\end{center}
}

 \author{
David S. Spiegel}

 \vspace{0.5\baselineskip}
 
\email{
dsp@astro.princeton.edu
}

\begin{abstract}
Although our solar system features predominantly circular orbits, the
exoplanets discovered so far indicate that this is the exception
rather than the rule.  This could have crucial consequences for
exoplanet climates, both because eccentric terrestrial exoplanets
could have extreme seasonal variations, and because giant planets on
eccentric orbits could excite Milankovitch-like variations of a
potentially habitable terrestrial planet’s eccentricity, on timescales
of thousands-to-millions of years.  A particularly interesting
implication concerns the fact that the Earth is thought to have gone
through at least one globally frozen, ``snowball'' state in the last
billion years that it presumably exited after several million years of
buildup of greenhouse gases when the ice-cover shut off the
carbonate-silicate cycle. Water-rich extrasolar terrestrial planets
with the capacity to host life might be at risk of falling into
similar snowball states.  Here we show that if a terrestrial planet
has a giant companion on a sufficiently eccentric orbit, it can
undergo Milankovitch-like oscillations of eccentricity of great enough
magnitude to melt out of a snowball state.
\end{abstract}

\section*{Proceeding}
Even very mild astronomical forcings can have dramatic influence on
the Earth's climate.  Although the orbital eccentricity varies between
$\sim$0 and only $\sim$0.06, and the axial tilt, or obliquity, between
$\sim$22.1$\degr$ and 24.5$\degr$, these slight quasi-periodic changes
are sufficient to help drive the Earth into ice ages at regular
intervals. Milankovitch articulated this possibility in his
astronomical theory of climate change.  Specifically, Milankovitch
posited a causal connection between three astronomical cycles
(precession -- 23~kyr period, and variation of both obliquity and
eccentricity -- 41-kyr and 100-kyr periods, respectively) and the
onset of glaciation/deglaciation.  Though much remains to be
discovered about these cycles, often in the literature referred to as
``Milankovitch cycles,''\footnote{Or as ``Croll-Milankovitch cycles''
  \cite{croll1875, milankovitch1941}.} they are now generally
acknowledged to have been the dominant factor governing the climate
changes of the last several million years \cite{berger1975,
  hays_et_al1976, berger1976, berger1978, berger_et_al2005}.

The nonzero (but, at just 0.05, nearly zero) eccentricity of Jupiter's
orbit is the primary driver of the Earth's eccentricity Milankovitch
cycle.  Were Jupiter's eccentricity greater, it would drive larger
amplitude variations of the Earth's eccentricity.  This same mechanism
might be operating in other solar systems.  In the last 15 years,
roughly $\sim$500 extrasolar planets have been discovered around other
stars, where an object is here defined as a ``planet'' by the
condition that it will not burn significant amounts of deuterium,
which corresponds approximately to 13 Jupiter masses
\cite{spiegel_et_al2010c}.  (This might not be the best way to define
exoplanets, but it is probably the most widely used.).  Among these
$\sim$500, there are many that have masses comparable to Jupiter's and
that are on highly eccentric orbits; $\sim$20\% of the known
exoplanets have eccentricities greater than 0.4, including such
extreme values as 0.93 and 0.97 (HD~20782b; HD~80606).  Furthermore,
tantalizing evidence suggests that lower mass terrestrial planets
might be even more numerous than the giant planets that are easier to
detect.  Therefore, it seems highly likely that many terrestrial
planets in our galaxy experience exaggerated versions of the Earth's
eccentricity Milankovitch cycle.

These kinds of cycles could have dramatic influence on life that
requires liquid water.  Since the seminal work of Milankovitch several
decades ago, a variety of theoretical investigations have examined the
possible climatic habitability of terrestrial exoplanets.  Kasting and
collaborators emphasized that the habitability of an exoplanet depends
on the properties of the host star \cite{kasting_et_al1993}.  Several
authors have considered how a planet's climatic habitability depends
on the properties of the planet, as well.  In particular, two recent
papers have focused on the climatic effect of orbital eccentricity.
Williams \& Pollard used a general circulation climate model to
address the question of how the Earth's climate would be affected by a
more eccentric orbit \cite{williams+pollard2003}.  Dressing et
al. used an energy balance climate model \cite{dressing_et_al2010} to
explore the combined influences of eccentricity and obliquity on the
climates of terrestrial exoplanets with generic surface geography (see
also \cite{williams+kasting1997} and \cite{spiegel_et_al2008,
  spiegel_et_al2009} for further description of the model).  A more
eccentric orbit both accentuates the difference between stellar
irradiation at periastron and at apoastron, and increases the annually
averaged irradiation.  Thus, periodic oscillations of eccentricity
will cause concomitant oscillations of both the degree of seasonal
extremes and of the total amount of starlight incident on the planet
in each annual cycle.  Since these oscillations depend on
gravitational perturbations from other companion objects, the present
paper can be thought of as examining how a terrestrial planet's
climatic habitability depends not just on its star, not just on its
own intrinsic properties, but also on the properties of the planetary
system in which it resides.

There is evidence that, at some point in the last billion years, Earth
went through a ``Snowball Earth'' state in which it was fully (or
almost fully) covered with snow and ice.  The high albedo of ice gives
rise to a positive feedback loop in which decreasing surface
temperatures lead to greater ice-cover and therefore to further net
cooling.  As a result, the existence of a low-temperature equilibrium
climate might be a generic feature of water-rich terrestrial planets,
and such planets might have a tendency to enter snowball states.  The
ice-albedo feedback makes it quite difficult for a planet to recover
from such a state.  In temperate conditions, the Earth's
carbonate-silicate weathering cycle acts as a ``chemical thermostat''
that tends to prevent surface temperatures from straying too far from
the freezing point of water.  A snowball state would interrupt this
cycle.  The standard explanation of how the Earth might have exited
its snowball state is that this interruption of the weathering cycle
would have allowed carbon dioxide to build up to concentrations
approaching $\sim$1 bar over a million-to-10-million years, at which
point the greenhouse effect would have been sufficient to melt the
ice-cover and restore temperate conditions.\footnote{Even an
  ice-encrusted planet in the habitable zone will eventually melt, due
  to the post-main sequence red-giant evolution of a Sun-like star, as
  the star grows larger and brighter.  The planet will not enjoy
  temperate conditions for long, however, as the continued growth in
  size and luminosity of the giant will eventually sterilize it of any
  water-based surface life.  Whether the Earth will be engulfed by the
  Sun in its giant phase (either by direct expansion or by tidal decay
  of its orbit), or will survive through the planetary nebula phase,
  remains an open question \cite{Nordhaus_and_Blackman_2006,
    nordhaus_et_al2010}. }

However, an exoplanet in a snowball state that is undergoing a large
excitation of its eccentricity might be able to melt out of its
globally frozen state in significantly less time, depending on the
magnitude of the eccentricity variations and on other properties of
the planet.  Exploring this possibility is the primary focus of
\cite{spiegel_et_al2010b}, in which, using an energy balance climate
model, we searched for orbital configurations that would lead to an
ice-covered planet melting out of the snowball state. In brief, we
found that orbital configurations that are not unlikely could cause a
snowball-Earth-analog to melt out by dint of increased eccentricity.

Figure 1 shows the temperature evolution of two cold-start planet
models, one of which (on the right) has a crude approximation of a
carbonate-silicate cycle incorporated in the infrared cooling term,
and the other (on the left) does not.  Both model planets have orbital
semimajor axis 1 AU, and are initialized to very cold
temperatures. The high orbital eccentricity of these models (0.8)
causes them to intercept more stellar irradiation over the annual
cycle than would a model on a circular orbit.  They therefore heat
rapidly and, with a crude accounting of the latent heat of
melting/freezing water \cite{spiegel_et_al2010b}, are eventually able
to melt through the ice layer.  Figure 2 shows two different
compressed Milankovitch-like cycles.  In each, a cycle that might take
10,000 -- 400,000 years is compressed to 25 years, for computational
feasibility and visualization purposes.  In one (the top row), the
planet is at semimajor axis 1 AU and has eccentricity varying
sinusoidally between 0 and 0.83.  In the other (bottom row), the
planet is at semimajor axis 0.8 AU and has eccentricity varying
between 0.1 and 0.33.  In each case, after several years, a
``catastrophic event'' dramatically increases the albedo for several
years, so as to plunge the model planet into a snowball state.  The
increasing eccentricity, then, eventually leads the planet to melt out
of the snowball state.  Finally, see Figures 3 and 4 of
\cite{spiegel_et_al2010b} for exemples of the magnitudes and
frequencies of Milankovitch-like eccentricity oscillations that can
result from gravitational interactions between an eccentric giant
planet and a terrestrial planet.  Though these kinds of oscillations
might be rare, they are not impossible.  Entirely prosaic planetary
system architectures can lead to less dramatic, but still highly
important, variations of a terrestrial planet's eccentricity.

In the coming years, as new observatories such as the James Webb Space
Telescope come online, exploring the atmospheres and atmospheric
dynamics of exoplanets will become an increasingly tractable research
problem.  Already, planets of the hot Jupiter class have been amenable
to investigation with the Spitzer Space Telescope, Kepler, and various
ground-based observatories (see, e.g., \cite{harrington_et_al2006,
  knutson_et_al2007b, fortney_et_al2008, spiegel_et_al2009b,
  showman_et_al2009, madhusudhan+seager2009, spiegel+burrows2010a},
and more).  It might even be possible to probe the atmospheric
composition of even extremely distant exoplanets, in the Galactic
bulge \cite{spiegel_et_al2005}.  Increasingly, it is possible to learn
about the properties of Neptune-mass exoplanets
\cite{demory_et_al2007, spiegel_et_al2010a, madhusudhan+seager2010}.
Discerning the spectral signatures of habitability and of life on
terrestrial planets will be the next frontier
\cite{kaltenegger_et_al2010}.  As the field of exoplanets matures, it
will be important to keep in mind that the long-term climatic
habitability of a planet might depend not just on the intrinsic
properties of the host star and of the planet itself, but also on the
detailed architecture of the planetary system in which the planet
resides.

\vspace{0.2in}

\begin{center}
 {\bf \sc Acknowledgments}
\end{center}
\vspace{-0.1in} {\small The author would like to thank the
  contributions of Sean Raymond, Courtney Dressing, Caleb Scharf,
  Kristen Menou, and Jonathan Mitchell.  Furthermore, DSS gratefully
  acknowledges the participants and organizers of the Milutin
  Milankovitch Anniversary Symposium, 2009, in particular Fedor
  Mesinger and Andre Berger.}

\bibliography{biblio_reduced.bib}
\newpage

\clearpage

\begin{figure}
\centerline{\plottwo{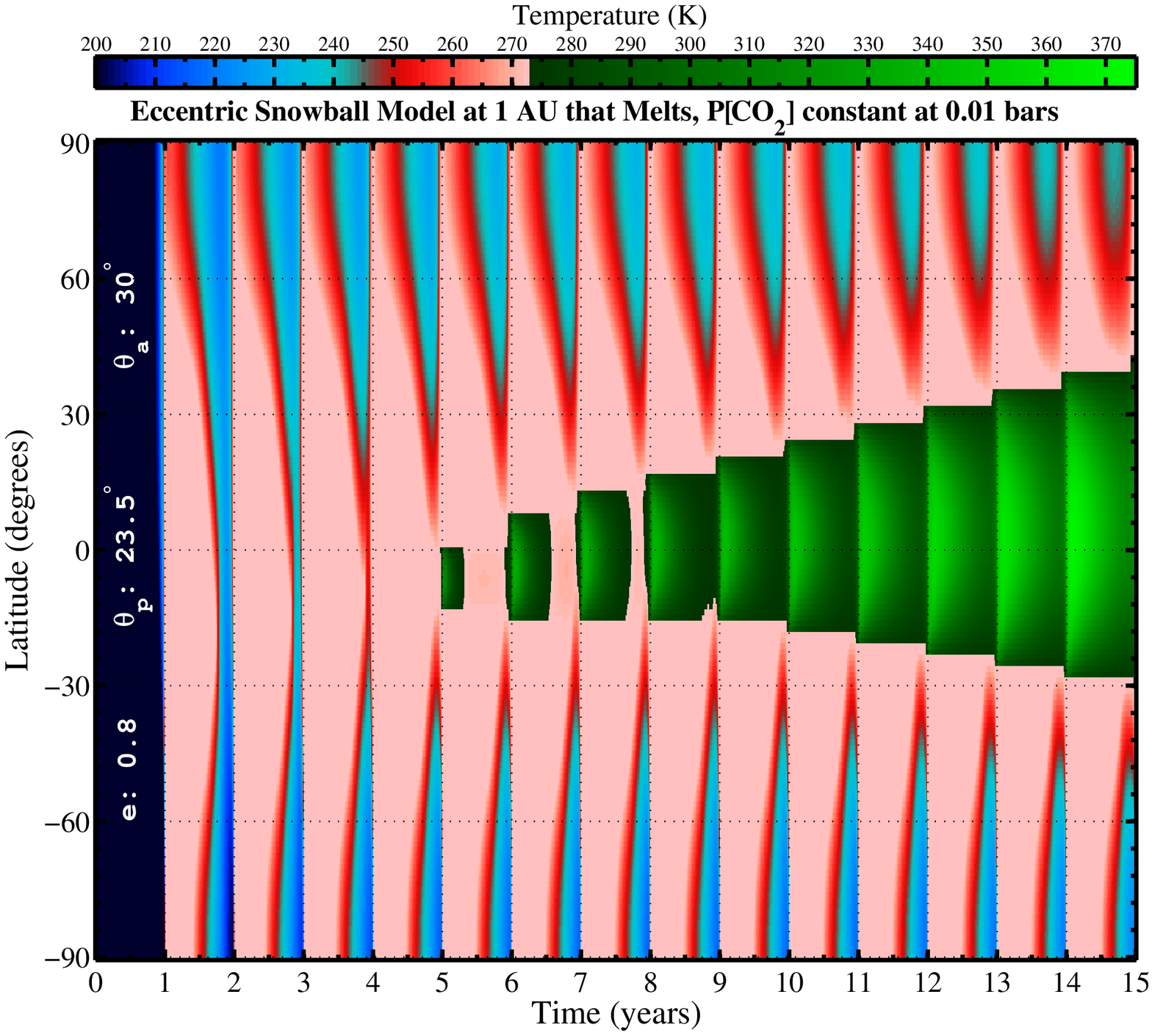}{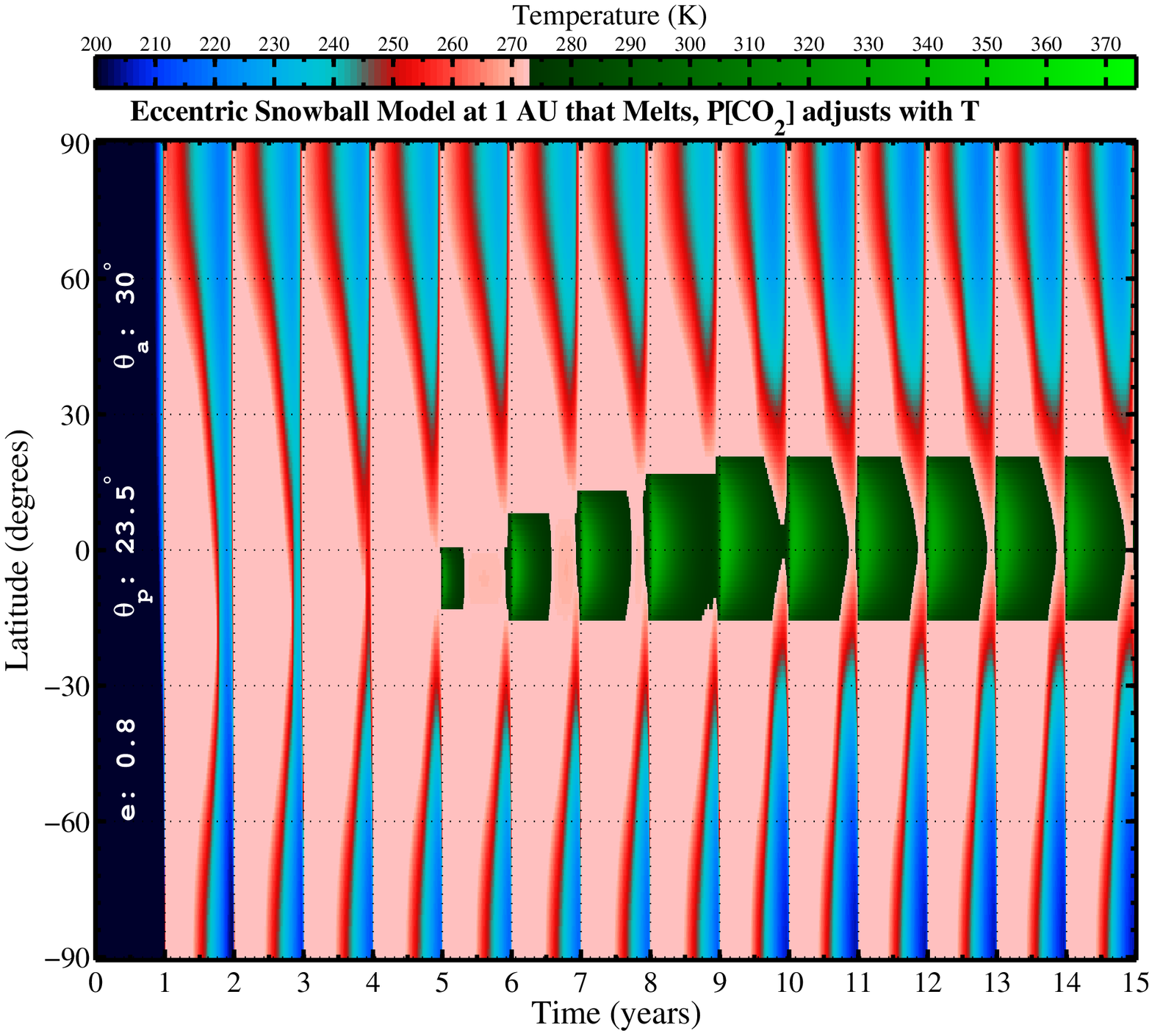}}
\caption{Temperature evolution maps for cold-start models at 1~AU.
  Both models have orbital eccentricity of 0.8 along with Earth-like
  $23.5\degr$ polar obliquity and 1~bar surface pressure. Temperature
  is initialized to 100~K, and quickly rises to near 273~K.  The
  melting of the ice-cover is handled in accordance with the
  prescription of \cite{spiegel_et_al2010b}.  {\bf Left:} CO$_2$
  partial pressure is held constant at 0.01~bars.  In this model, once
  the equatorial region melts, the region of surface that has melted
  ice-cover grows steadily until the entire planet has melted, and
  temperatures eventually grow to more than 400~K over much of the
  planet (not shown).  {\bf Right:} CO$_2$ partial pressure varies
  with temperature, in a crude simulation of a ``chemical
  thermostat''.  In this model, the climate reaches a stable state
  with equatorial melt regions and polar ice-cover.}
\label{fig:tempmap}
\end{figure}

\newpage
\clearpage

\begin{figure}
\centerline{\plottwo{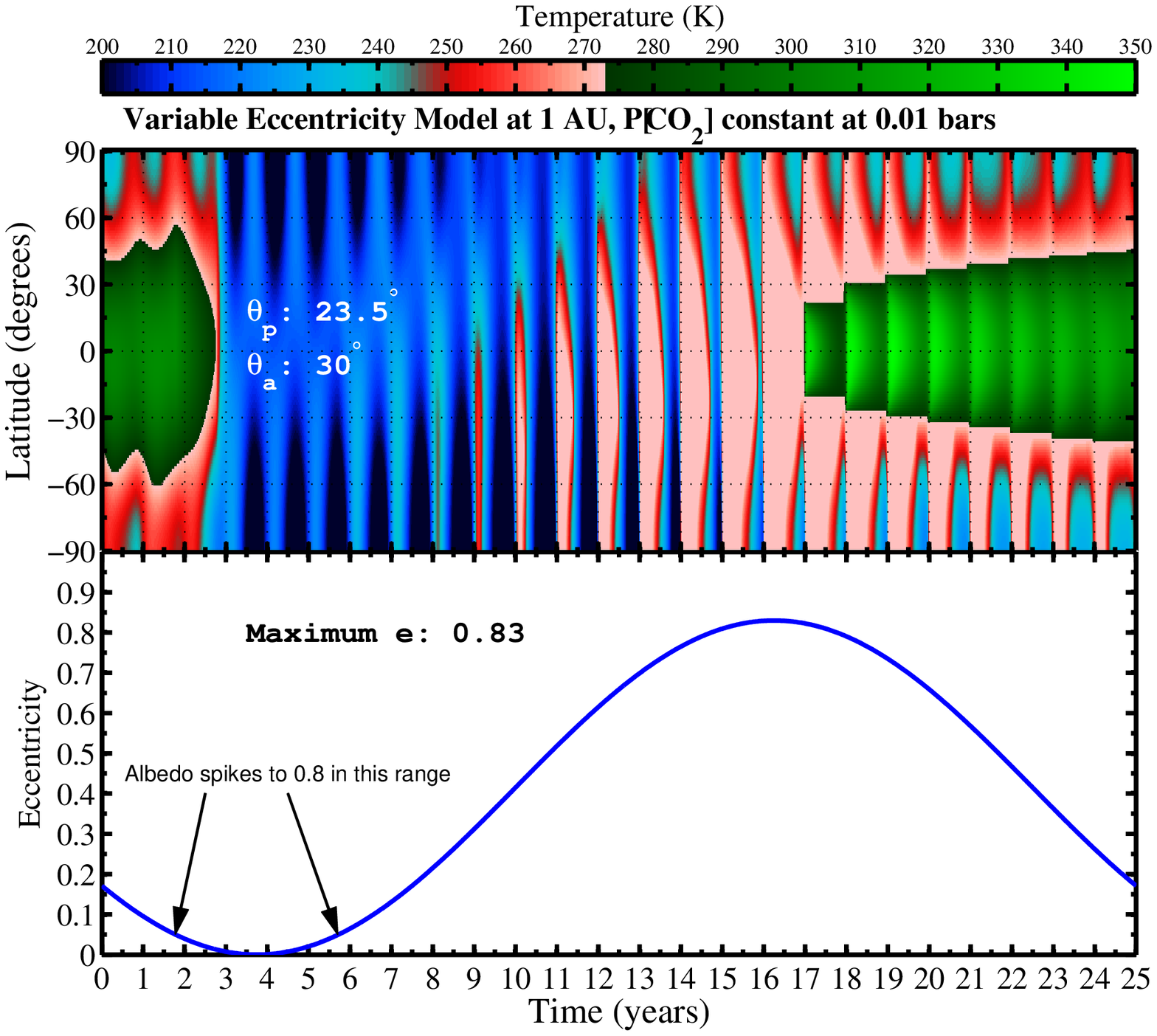}{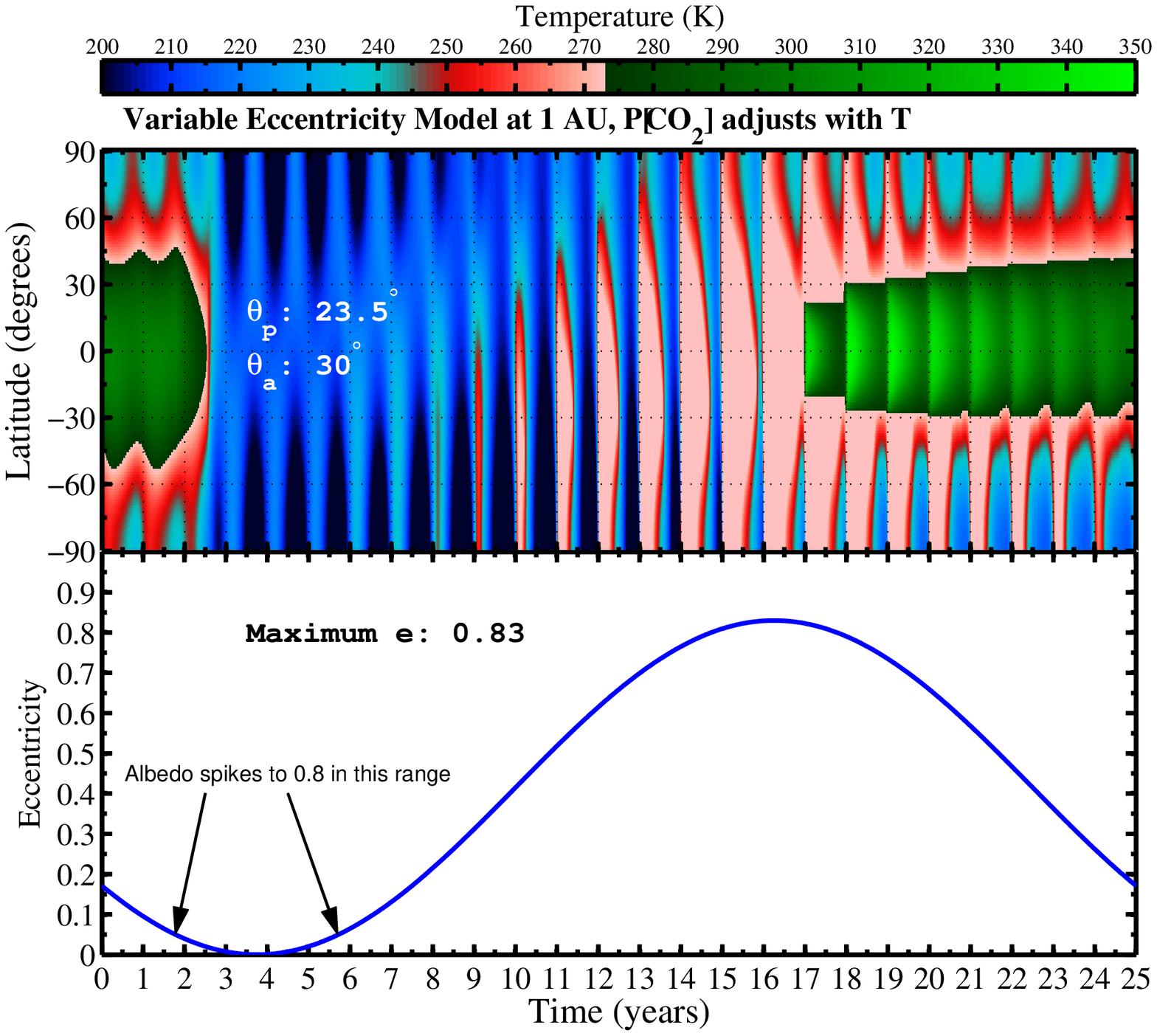}}
\centerline{\plottwo{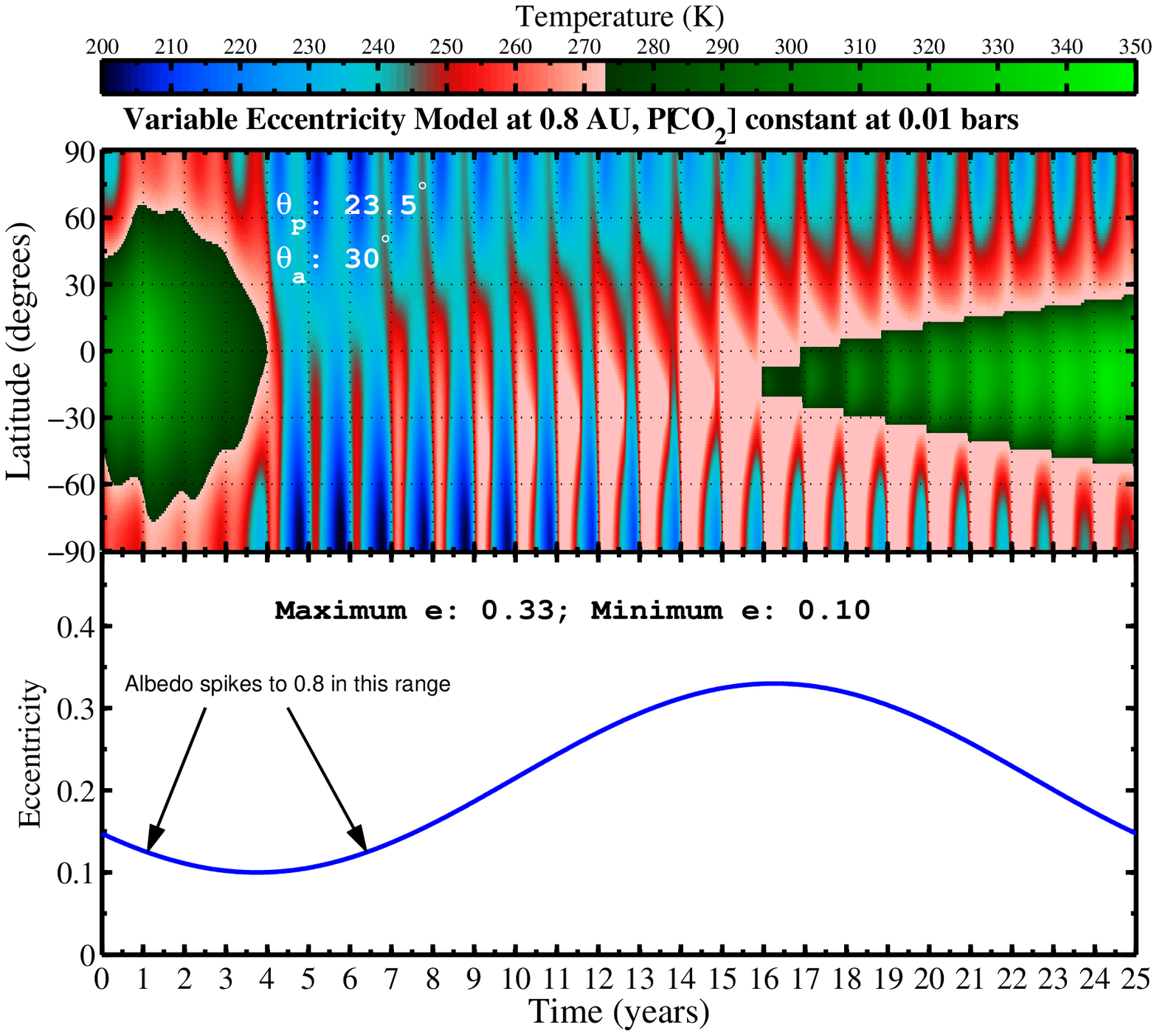}{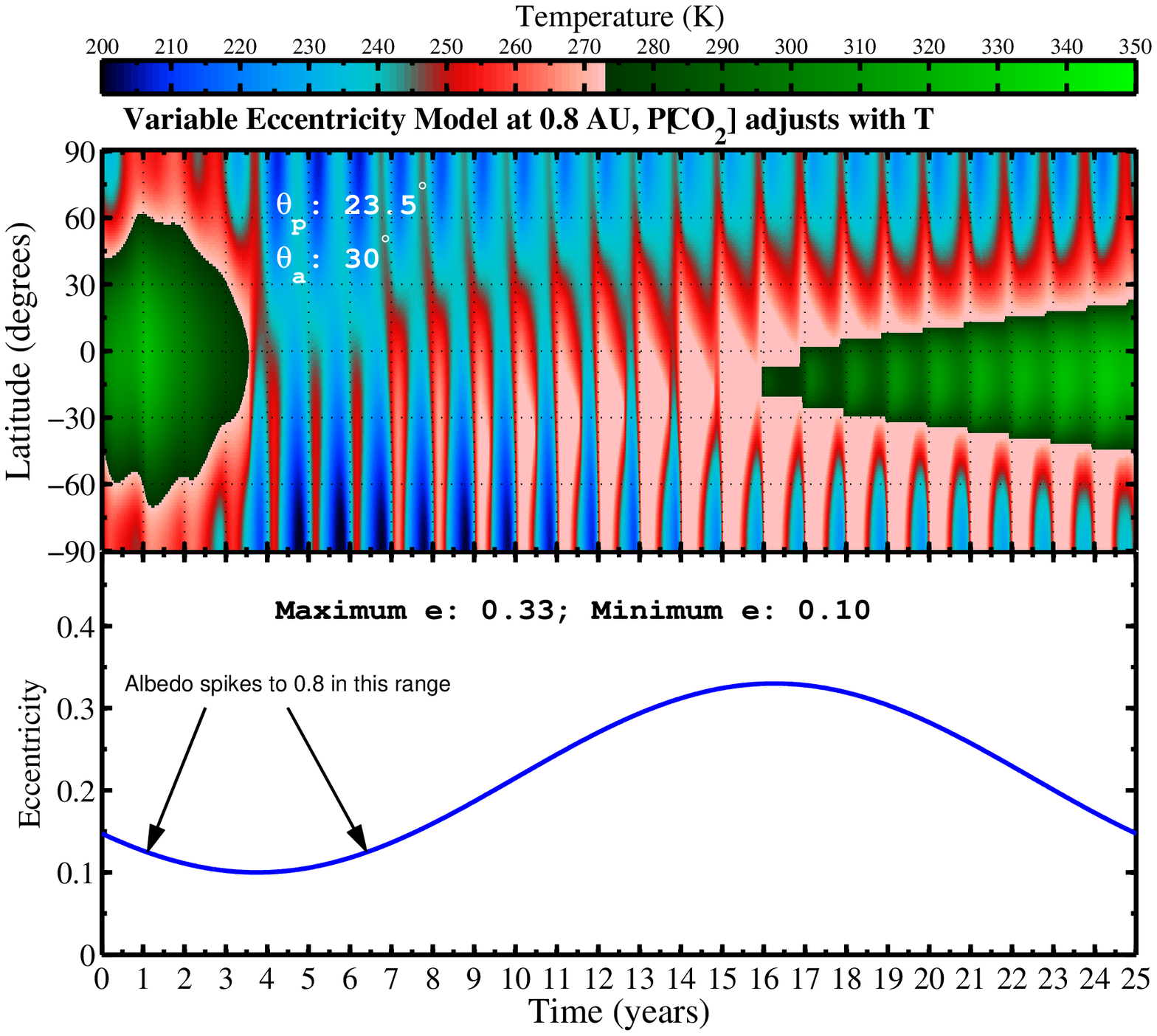}}
\caption{Compressed Milankovitch-like evolution of eccentricity and
  temperature at 1~AU and at 0.8~AU.  Planets are initialized with
  warm equator and cold poles, similar to present-day Earth.  In the
  top row (1~AU), the model planets are the same as in
  Fig.~\ref{fig:tempmap}, except the eccentricity varies sinusoidally
  between 0 and 0.83 with a 25-year period, to simulate a
  time-acceleration (by a factor of $\sim$10$^2$ to $\sim$10$^4$) of a
  Milankovitch-like cycle.  When the eccentricity falls below 0.05,
  the planet's albedo spikes to 0.8, simulating a catastrophic event
  that plunges the planet into a snowball state, with the latent heat
  prescription of \cite{spiegel_et_al2010b}. In the bottom row
  (0.8~AU), the eccentricity varies between 0.1 and 0.33, also with a
  25-year period.  {\bf Left:} CO$_2$ partial pressure is held fixed
  at 0.01~bars.  As in the left panel of Fig.~\ref{fig:tempmap}, these
  planets do not establish a temperate equilibrium.  {\bf Right:}
  CO$_2$ partial pressure varies with temperature.  Here, temperature
  increases are muted by reduced greenhouse effect once the ice-cover
  has melted somewhere.}
\label{fig:tempmap_eccen}
\end{figure}

\end{document}